\documentclass[a4paper]{article}

\usepackage{INTERSPEECH2018}

\newcommand{\rpm}{\sbox0{$1$}\sbox2{$\scriptstyle\pm$}
  \raise\dimexpr(\ht0-\ht2)/2\relax\box2 }

\usepackage{enumitem}
\usepackage[font=small,skip=1pt]{caption}
\title{The PRIORI Emotion Dataset: \\ Linking Mood to Emotion Detected In-the-Wild}
\name{Soheil Khorram$^{\star}$$^{\dagger}$, Mimansa Jaiswal$^{\star}$, John Gideon$^{\star}$, Melvin McInnis$^{\dagger}$, Emily Mower Provost$^{\star}$}
\address{Departments of: Computer Science and Engineering$^{\star}$ and Psychiatry$^{\dagger}$, University of Michigan}
\email{\{khorrams, mimansa, gideonjn, mmcinnis, emilykmp\}@umich.edu}

\begin{document}

\maketitle
\begin{abstract}
Bipolar Disorder is a chronic psychiatric illness characterized by pathological mood swings associated with severe disruptions in emotion regulation. Clinical monitoring of mood is key to the care of these dynamic and incapacitating mood states. Frequent and detailed monitoring improves clinical sensitivity to detect mood state changes, but typically requires costly and limited resources. Speech characteristics change during both depressed and manic states, suggesting automatic methods applied to the speech signal can be effectively used to monitor mood state changes.  However, speech is modulated by many factors, which renders mood state prediction challenging.  We hypothesize that emotion can be used as an intermediary step to improve mood state prediction.  This paper presents critical steps in developing this pipeline, including (1) a new \emph{in the wild} emotion dataset, the PRIORI Emotion Dataset, collected from everyday smartphone conversational speech recordings, (2) activation/valence emotion recognition baselines on this dataset (PCC of 0.71 and 0.41, respectively), and (3) significant correlation between predicted emotion and mood state for individuals with bipolar disorder. This provides evidence and a working baseline for the use of emotion as a meta-feature for mood state monitoring.
\end{abstract}

\section{Introduction}
Bipolar disorder (BD) is a severe, chronic mental illness that typically begins in early adulthood and is characterized by periodic and pathological mood changes ranging from extreme lows (depression) to extreme highs (mania) \cite{belmaker2004bipolar}. It is found in $1\%$ of the world's population, with a core clinical expression pattern related to emotion, energy, and psychomotor activity \cite{merikangas2011prevalence}. These core clinical signs and symptoms are monitored to gauge the health and progress of the individual in treatment  \cite{belmaker2004bipolar}. The dynamic nature of BD demands efficient clinical monitoring to detect mood changes in sufficient time to treat or mitigate their severity. Intense clinical monitoring is effective but unrealistic due to cost and the availability of skilled health care providers \cite{morriss2007interventions}. Automatic passive mood monitoring addresses the need for ongoing monitoring in a cost efficient manner to predict the course and outcome of chronic human disease such as BD. 

Current mood recognition systems are focused on mapping between speech to mood directly, which is challenging due to the complexity of the speech signal. We hypothesize that emotion can simplify mood prediction by acting as an intermediary between speech (rapidly varying) and mood (slowly varying).  Further, one of the hallmark symptoms of BD is emotion dysregulation, suggesting that the tracking of emotion changes will provide important insights into an individual's mood variation. In this paper, we define emotion in terms of valence (positive vs. negative) and activation (calm vs. excited), both of which are observable from expressed behaviors such as speech. 

Strategies for mobile monitoring for mental health have mostly relied upon self-reported diagnosis of a disorder on a device or social media to identify features indicative of the disorder \cite{berrouiguet2018ehealth}, be it anxiety \cite{kumar2018mobile}, BD \cite{nicholas2017reviews}, depression or anorexia \cite{frank2018smart}. However, these interactive self-reports are often incomplete or misleading \cite{neary2018state}. 
An alternative approach is to directly recognize mood from observed behavior.  Vanello et al.~\cite{vanello2012speech} and Faurholt et al.~\cite{faurholt2016voice} investigated how speech features could be used to characterize mood states. Muaremi et al. in~\cite{muaremi2014assessing} used statistics of phone calls, such as duration and frequency, to predict mood episodes.  Our own work has demonstrated that properties such as speaking rate are also effective for detecting mood~\cite{khorram2016recognition, gideon2016mood}. However, a recurring theme in these studies is the challenge associated with detecting mood directly from speech, due in part to the highly varying nature of the speech signal. We hypothesize that we will be able to improve mood detection by using an intermediary (emotion) that has more slowly varying properties.

The relationship between emotion and mood has been gaining attention. Stasak et al. investigated the utility of using emotion to detect depressed speech \cite{stasak2016investigation}, using the AVEC 2014 dataset \cite{valstar2014avec}.  However, these data were collected in a controlled environment, potentially limiting their use ``in the wild''. Carrillo et al. identified a relationship between emotional intensity and mood in the context of BD \cite{carrillo2016emotional}. However, they relied upon transcribed interviews, rather than on acoustics directly.  

The work presented in this paper leverages the PRIORI (PRedicting Individual Outcomes for Rapid Intervention) dataset, a longitudinal dataset of natural speech patterns from individuals with BD~\cite{karam2014ecologically}. Data were collected from individuals with BD for six to twelve months using smartphones with a secure app that recorded their side of all phone conversations. They were assessed weekly for depressive and manic symptoms using standardized scales by a study clinician \cite{khorram2016recognition,gideon2016mood,karam2014ecologically}. We analyze a subset of this dataset, referred to as the PRIORI Emotion Dataset, which is annotated with labels of valence and activation. This provides an opportunity to associate natural expressions of emotion with changes in mood.


In this paper, we address: (1) the predictability of emotion in natural smartphone conversations and (2) the relationship between mood and natural expressions of emotion. We describe the collection of the full PRIORI dataset, as well as the creation of the PRIORI Emotion Dataset. We establish natural speech emotion classification baselines on this dataset, which achieve a Pearson correlation coefficient (PCC) of 0.71 and 0.41 for detecting activation and valence, respectively. Finally, we demonstrate that there is a significant positive correlation between heightened mood and both activation and valence. Critically, we note that these emotion patterns are inherently subject-dependent, highlighting the importance of attuning to individual variability when designing mental health monitoring support.

\begin{table}[t]
\center                                         
\begin{tabular}{c|cccc}
\centering
Mood       & HamD & YMRS & Number & \# Per Subject \\ \hline
Euthymic   & $\leq$6 & $\leq$6  & 70          & 5.8 $\rpm$ 3.4            \\
Manic      & $<$10 & $\geq$10   & 27          & 2.7 $\rpm$ 1.9            \\     
Depressed  & $\geq$10 & $<$10   & 120         & 10.0 $\rpm$ 6.4           \\
Excluded   & Else & Else        & 96          & 8.7 $\rpm$ 6.7
\end{tabular}
\caption{Mood state categories defined by HamD and YMRS measures, including the number of total assessments and the mean and standard deviation of assessments per subject.\vspace*{-5mm}}
\label{AssessAmountTable}
\end{table}

\section{PRIORI Dataset}
\label{sec:priori}
The PRIORI dataset is composed of one-sided natural conversations recorded during daily smartphone usage (Samsung Galaxy S3, S4, S5) \cite{karam2014ecologically,gideon2016mood,khorram2016recognition}. The participants include 51 individuals with BD and nine healthy controls.  The inclusion criteria were: BD type I or II, no medical or neurological disease, and no active history of substance abuse. All study participants were provided with a smartphone and were asked to use the smartphone as their primary device. The smartphone has an app that runs silently in the background, recording the speech with 8 kHz sampling frequency, and uploading the recordings to our servers for analysis. Participants were enrolled for an average of 32$\rpm$16 weeks. The collection includes 52,931 calls and over 4,000 hours of speech.

Participants were clinically evaluated in weekly \emph{assessment calls} to assess the level of depression (Hamilton Depression Scale (HamD) \cite{hamilton1976hamilton}) and mania (Young Mania Rating Scale (YMRS) \cite{young1978rating}). All other recordings are referred to as \emph{personal calls}. We assign mood labels to all assessment calls based on the HamD and YMRS scales. We define four mood labels: euthymic, manic, depressed, and excluded. A call is labelled \emph{euthymic} if it has a score of six or less on both the HamD and YMRS scales; \emph{manic} if the score is ten or greater on the YMRS and less than ten on the HamD; and \emph{depressed} if the score is ten or greater on the HamD and less than ten on the YMRS. All other assessments are excluded from our experiments (Table~\ref{AssessAmountTable}).

\vspace{-2mm}
\section{PRIORI Emotion Annotation}
This work explores the relationship between mood state and emotion expression, which necessitates access to a labeled corpus over which emotion can be detected and emotion classification algorithms can be validated.  However, there are no natural smartphone conversational speech datasets annotated in this manner.  We addressed this limitation by generating the PRIORI Emotion Dataset, a subset of the larger PRIORI dataset. The PRIORI Emotion Dataset contains manual valence/activation annotations of both assessment and personal calls. We use a dimensional labeling strategy in this work \cite{bradley1994measuring}, motivated by the concept of \emph{core affect}~\cite{russell2003core}. This construct provides a de-contextualized manner of considering emotion expression.

The PRIORI Emotion dataset includes natural conversational speech from 12 subjects, seven females and five males, totaling 11,337 calls (928 hours). The selected subjects are between 24 and 63 years old. We selected the subjects based on three factors: (1) BD diagnosis, which allows us to examine the link between emotion and bipolar mood (future work will focus on healthy controls); (2) used Samsung S5, which provides microphone consistency, lack of which was identified as a challenge in our prior work \cite{gideon2016mood}; (3) provided informed consent for annotation of personal calls, which allows us to generate ground-truth emotion labels. 

We then annotate a subset of these data using: (1) segmentation, (2) segment selection, (3) segment inspection, and (4) segment annotation. We explain each in the following sections.

\textbf{Segmentation:}
We first filter the set of calls to exclude all recordings longer than one hour. This restriction is due to the large memory requirements and processing time associated with these data. We then perform speech activity detection (SAD), using the COMBO-SAD algorithm introduced by Sadjadi and Hansen \cite{sadjadi2013unsupervised}. We form contiguous segments following the methodology used in our prior work \cite{gideon2016mood}. The resulting segments contain continuous speech with no intermediate silence. This procedure provides 167,339 segments (10,563 segments from assessment calls and 156,776 segments from personal calls) with the average length of 6.32\rpm5.89 seconds.

\textbf{Segment Selection:}
We identified a subset of segments for manual annotation from the assessment and personal calls. Our first filter was for segment length, to increase the likelihood that segments contained sufficient data to assess, but were not so long that the emotion would vary over the course of the segment.  We exclude segments shorter than three seconds and longer than 30 seconds.  Next, we sampled from both personal calls and assessment calls.  Assessment calls are important because they are the only calls that are directly associated with mood labels.  Personal calls are important because they contain natural unstructured speech.  Therefore, we sampled from both to ensure a diversity of examples.  For each assessment call, we select up to ten random segments. For personal calls, we sample as a function of proximity in time to assessment calls, preferring those that occurred closer to the assessments.  We select 1,200 segments randomly considering the weight of $max(4-d, 1)$, where d is the number of days between the call and its future assessment day. Calls on the day of assessment receive a weight of four (these are most closely linked to the HamD/YMRS score).  Other calls receive a weight that reduces linearly up to 3 days before assessment day, calls outside this range have a weight of one.  This results in 17,237 segments, 2,837 and 14,400 segments from the assessment and personal calls, respectively.


\textbf{Segment Inspection:}
We manually inspected each segment prior to annotation and removed those that were deemed inappropriate for the annotation task, if: (1) background noise dominates the speech signal, (2) speech content of the segment lasts less than two seconds, (3) subject is not talking to the phone (e.g., talking to someone else in the room), (4) emotion clearly varies over the course of the segment, and (5) segment contains identifiable information (e.g., name, address, phone number, etc.).  This results in \textbf{13,611} segments (\textbf{25.20} hours), 2,209 and 11,402 segments from the assessment and personal calls, respectively.

\textbf{Segment Annotation:}
We annotated the activation and valence of the 13,611 speech segments using the established pictorial manikins method across a 9-point Likert scale (1: very low to 9: very high) \cite{bradley1994measuring}. There were 11 annotators  (7 female, 4 male) aged between 21 and 34 
and native speakers of English.


We conducted a training session for each annotator, including a training video and manuscript, to introduce the annotation software and provide annotation examples. In the training session, annotators were asked to consider two important points: \vspace{-15pt}
\begin{enumerate}[leftmargin=*]
\item Although challenging, we asked that annotators to only consider the acoustic characteristics of the recordings, not the lexical content. They were asked to avoid letting speech content ``color'' their activation and valence labels. \vspace{-3pt}
\item We asked that annotators consider the subject-specificity of emotion expression.  When approaching a new subject, annotators were asked to spend some time listening to a few segments without assigning a rating in order to get a better sense of what that person's baseline sounds like.
\end{enumerate}
\vspace{-3pt}

We further supported the assessment of subject-dependent emotion patterns by providing \emph{individual context} for each participant.  The annotation software randomly selected a participant and presented all segments of that participant, in random order, to the annotator before moving on to the next participant's segments.  In this way, annotators can consider participant-specific features to define emotion labels more accurately. 

\begin{figure}[t]
    \centering
    \def\factor{1}
    \includegraphics[width=\factor\linewidth]
    {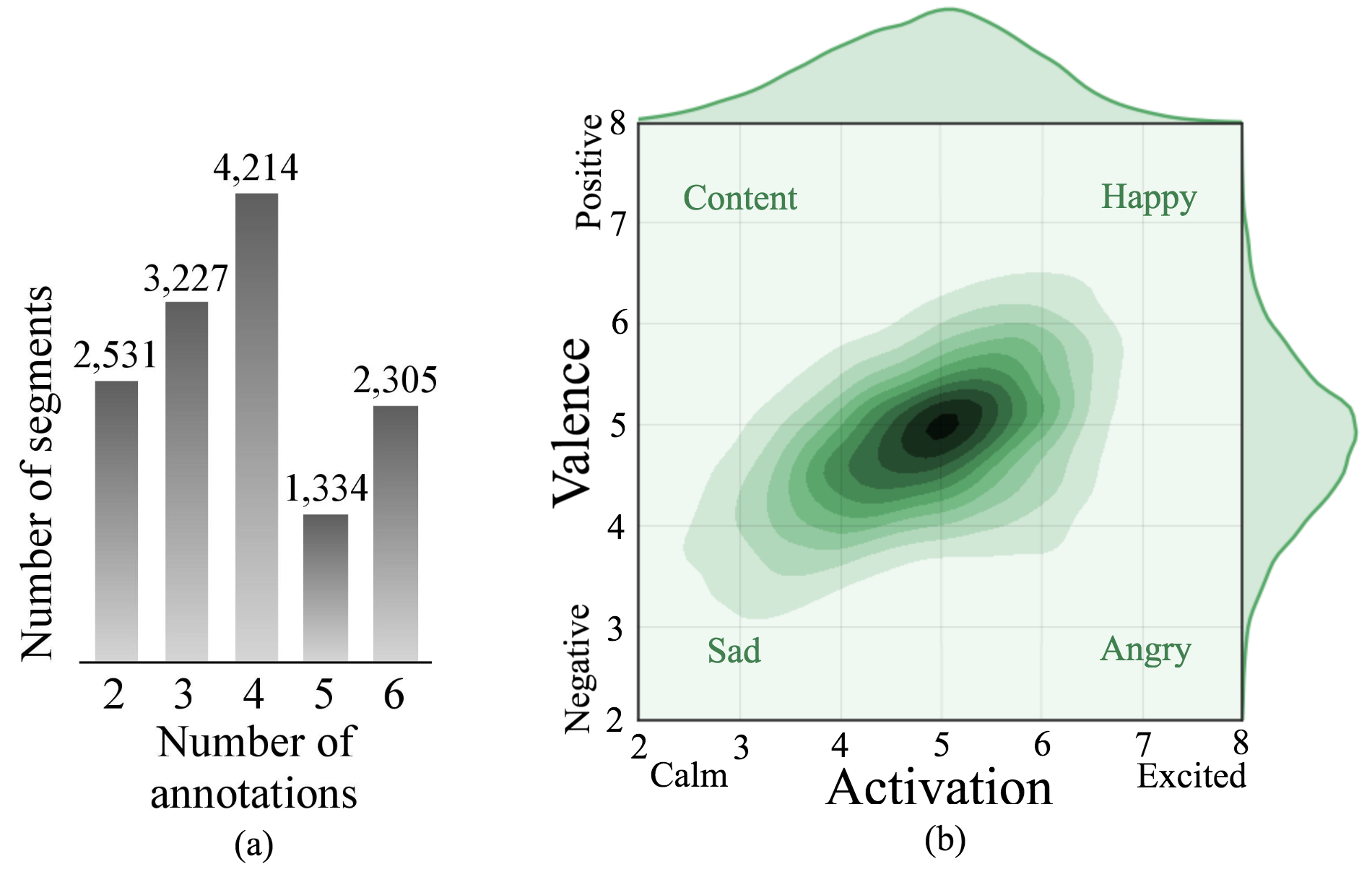}
    \caption{(a) Distribution of the number of labels annotated for the segments. (b) Distribution of the activation and valence ratings in the PRIORI Emotion Dataset. Categorical labels are provided only as reference points for the four quadrants.\vspace*{-7mm}}
    \label{DatasetHist}
\end{figure}

We collected between two and six labels for each segment (\textbf{3.83 $\pm$ 1.31} labels per segment). Figure \ref{DatasetHist} shows the distribution of the number of annotations for each segment. See Figure \ref{DatasetHist} for a distribution of the activation and valence labels defined by the annotators. We found that the activation and valence values are significantly correlated with a PCC of 0.46 ($p < 0.01$).

\vspace*{-1mm}
\section{Methods}
We describe two emotion prediction systems used in this work. The first system is a deep feed-forward neural network (FFNN) that operates on the eGeMAPS feature set~\cite{eyben2016geneva}. The second system is a convolutional neural network (CNN) classifier that operates on log Mel-frequency bank (log-MFB) features~\cite{busso2007using}.\vspace{-2mm}

\subsection{Acoustic Features}

\textit{\textbf{eGeMAPS --}} 
The eGeMAPS feature set is a carefully designed standardized acoustic feature set for affective computing. It is an 88-dimensional feature vector that includes features relating to energy, excitation, spectral, cepstral, and dynamic information. 
We extract the eGeMAPS feature set using the openSMILE toolkit with default parameters~\cite{eyben2010opensmile}, as in our previous work~\cite{gideon2017progressive}.
\\
\textit{\textbf{Log-MFB --}} 
Previous research has demonstrated that log mel-frequency bank (log-MFB) spectral features outperform other temporal frame-level acoustic features such as MFCCs~\cite{busso2007using,khorram2017capturing,aldeneh2017pooling}. The details of our log-MFB extractor are explained in~\cite{khorram2017capturing}. We extract 40-dimensional log-MFB features with $25$ms frame length and $10$ms frame shift using the Kaldi toolkit~\cite{povey2011kaldi}. We perform global $z$-normalization on all features.
\\
\textit{\textbf{Feature Handling --}} 
Unlike eGeMAPS that is a fixed-length feature vector for each recording, log-MFB has a variable length (its length is proportional to the length of the recording). Training a neural network with a variable-length input is problematic; the network structure must be consistent with the input length while the length varies for different recordings. We use the zero-padding technique to avoid this problem.  

\begin{table}[t]
\center                                         
\begin{tabular}{c|cc}
Measure & eGeMAPS FFNN & MFBs Conv-Pool \\ \hline
Activation PCC & 0.642 $\rpm$ 0.076 & \textbf{0.712$\rpm$0.077} \\
Activation CCC & 0.593 $\rpm$ 0.071 & \textbf{0.660$\rpm$0.090} \\
Activation RMSE & 0.207 $\rpm$ 0.012 & 0.201 $\rpm$ 0.028 \\
\rule{0pt}{3ex} Valence PCC & 0.271 $\rpm$ 0.053 & \textbf{0.405$\rpm$0.062} \\
Valence CCC & 0.191 $\rpm$ 0.031 & \textbf{0.326$\rpm$0.052} \\
Valence RMSE & 0.199 $\rpm$ 0.015 & 0.194 $\rpm$ 0.016 \\
\end{tabular}
\caption{The results of the emotion prediction systems. A bolded result indicates that the conv-pool method is significantly better than FFNN, using a paired t-test with $p < 0.01$.\vspace*{-4mm}}
\label{EmotionResults}
\end{table}

\begin{table*}[t]
\center              
\setlength\tabcolsep{5pt}
\begin{tabular}{c|cccccccccccc}
Subject & 1 & 2 & 3 & 4 & 5 & 6 & 7 & 8 & 9 & 10 & 11 & 12 \\ \hline
Mean Activation (Mania) & \textbf{-0.12} & \textbf{1.06} & \textbf{1.56} & \textbf{0.45} & -0.13 & \textbf{-0.03} & \textbf{0.73} & 0.02 & \textbf{1.35} & -0.86 & - & - \\
Mean Activation (Depression) & -0.49 & -0.81 & 0.83 & -0.18 & -0.20 & -0.34 & 0.29 & -0.26 & -1.56 & -0.35 & -0.61 & -3.33 \\
\rule{0pt}{3ex} Mean Valence (Mania) & \textbf{0.05} & \textbf{0.72} & \textbf{0.66} & \textbf{0.35} & -0.25 & \textbf{0.32} & \textbf{0.89} & -0.02 & \textbf{0.55} & -0.60 & - & - \\
Mean Valence (Depression) & -0.42 & -0.98 & 0.13 & -0.08 & -0.31 & -0.37 & 0.37 & -0.27 & -1.39 & 0.05 & 0.00 & -3.13
\end{tabular}
\caption{Subject-specific mean activation and valence ratings for manic and depressed states. Dashes indicate that the subject did not have any manic episodes. Bold font indicates statistically significant differences between manic and depressed states  (t-test, $p < 0.01$).\vspace*{-3.5mm}}
\label{MoodResults}
\end{table*}

\begin{table*}[t]
\center              
\setlength\tabcolsep{5pt}
\begin{tabular}{c|cccccccccccc}
Subject & 1 & 2 & 3 & 4 & 5 & 6 & 7 & 8 & 9 & 10 & 11 & 12 \\ \hline
PCC of Activation and YMRS & -0.02 & \textbf{0.46} & \textbf{0.19} & 0.04 & \textbf{0.11} & 0.06 & \textbf{0.13} & \textbf{0.19} & 0.08 & -0.18 & \textbf{0.39} & \textbf{0.55} \\
PCC of Activation and HamD & \textbf{-0.11} & \textbf{-0.13} & \textbf{-0.25} & \textbf{-0.11} & -0.04 & -0.10 & 0.04 & \textbf{-0.10} & \textbf{-0.57} & -0.11 & \textbf{-0.16} & \textbf{-0.80} \\
\rule{0pt}{3ex} PCC of Valence and YMRS  & 0.00 & \textbf{0.44} & \textbf{0.17} & 0.00 & \textbf{0.08} & \textbf{0.13} & \textbf{0.19} & \textbf{0.20} & -0.04 & \textbf{-0.14} & \textbf{0.40} & \textbf{0.55} \\
PCC of Valence and HamD & -0.10 & \textbf{-0.13} & \textbf{-0.33} & \textbf{-0.10} & \textbf{-0.06} & \textbf{-0.13} & \textbf{0.09} & \textbf{-0.10} & \textbf{-0.52} & -0.05 & 0.09 & \textbf{-0.76} \\
\end{tabular}
\caption{The PCC between each emotion rating and mood state. Even subjects without manic episodes are included in this analysis, as YMRS ratings are available. Significant correlations are bolded (t-test, $p < 0.01$).\vspace*{-5mm}}
\label{MoodResults_pcc}
\end{table*}

\subsection{Models}
\textit{\textbf{Deep FFNN --}}
This network contains a stack of fully-connected dense layers with tanh activation functions followed by an output layer with a linear activation function. We predict activation and valence values independently. 
\\
\textit{\textbf{Conv-Pool --}}
We implemented Conv-Pool network, proposed in~\cite{aldeneh2017using}, due to its high emotion recognition accuracy on the IEMOCAP~\cite{busso2008iemocap} and MSP-IMPROV~\cite{busso2017msp} datasets. The Conv-Pool network contains three major components: (1) a stack of convolutional layers; (2) a global pooling over time layer; and (3) a stack of dense layers. The convolutional layers create a sequence of feature maps that identify emotionally salient regions within variable length utterances. The global pooling layer automatically extracts a set of call-level statistics. Aldeneh et al. found that a max-pooling layer is effective for emotion recognition~\cite{aldeneh2017using}. Finally, the stack of dense layers predicts the labels from the call-level features. We used ReLU and linear activation functions for intermediate and output layers.

\section{Emotion Detection Baselines}
The ground truth annotation for a segment is the average of all individual annotations. We normalized the ground truth labels by subtracting the rating midpoint of 5 and scaling to the range of $[-1,1]$. Let x be a rating, the normalization was performed through $\frac{x-5}{4}$. Our preliminary analyses showed that this transformation helped the networks to learn the bias and standard deviation of the labels more quickly. 

We will measure system performance using the repeated cross-validation method introduced in~\cite{bouckaert2004evaluating}. Each experiment is repeated for five total runs, where a run is defined as six randomly selected folds. In each run, the folds are shuffled by randomly assigning two subjects to each of the six folds. We then use round-robin cross-validation: at each step, one fold (two subjects) is assigned to testing, one is used for tuning parameters and early stopping, and the rest are used for training. This procedure generates one test measure per fold, resulting in six measures. Over the course of the five runs, a matrix of 6-by-5 test measures is output. We report the mean over all experiments as the experiment mean. The experiment standard deviation is the mean standard deviation within runs. Significance was determined using a repeated cross-validation paired $t$-test with six degrees of freedom, as shown in \cite{bouckaert2004evaluating}.

We implemented both networks using Keras with TensorFlow backend~\cite{chollet2015keras}. We optimized RMSE during training through the Adam optimizer~\cite{chollet2015keras} with a fixed learning rate of 0.0001. All weights were initialized using the Xavier uniform algorithm~\cite{chollet2015keras} and all bias parameters were set to zero. We set epoch size to 64. To train the FFNN, we performed cross-validation to tune the number of dense layers (2,4,8) and the number of nodes in each layer (200, 400, 800). To train Conv-Pool network, we set the number of initial convolutional layers and final dense layers equal and validated them over a set of (2,3). The number of nodes (200,400) and the length of the convolution kernels (4,8) were also validated. We trained FFNN and Conv-Pool networks for 100 and 15, epochs respectively. For each test fold, we selected the best epoch and the best network structure based on the validation concordance correlation coefficient (CCC) value. 

We use three popular metrics to compare the resulting networks: PCC, CCC, and RMSE. Using Conv-Pool, we achieved a PCC of 0.712 and 0.405 for activation and valence, and found that Conv-Pool has significantly better performance than FFNN using all measures except RMSE (Table \ref{EmotionResults}). This supports previous work that demonstrated the importance of modeling temporal characteristics of speech in emotion recognition~\cite{aldeneh2017using}. We hypothesize that the lack of improvement for RMSE is due to the fact that RMSE places more weight on selecting the correct bias of the ratings. Subject-dependent or speaker-adapted models may improve RMSE. Finally, as shown in previous datasets, activation is easier to predict from speech than valence~\cite{truong2009arousal}. \vspace*{-2mm}

\section{Mood Analysis}
In this section, the link between BD mood states and predicted emotion is tested. To facilitate this analysis, we use our Conv-Pool models to predict emotion on the 10,563 assessment call segments. We use the 30 different models from the repeated cross-validation as an ensemble and take the mean output. 

We normalize the predicted emotion labels using subject-dependent euthymic z-normalization. Our preliminary analyses demonstrated the importance of considering how a subject varies about his/her \emph{own} baseline, which is defined as his/her euthymic periods.  We calculate the mean and standard deviation of the valence/activation ratings over all calls associated with euthymic mood states. We then normalize each segment based on these values, reducing the effect of subject biases.

Table \ref{MoodResults} shows the mean activation and valence ratings, calculated over the segments in each of the different mood states. Table \ref{MoodResults_pcc} shows the PCC between each of the dimensional emotion ratings (activation and valence) and each of the mood ratings (YMRS and HamD). We note that:
\begin{itemize}[leftmargin=*]
    \vspace{-0.5mm}
    \item In the majority of subjects, the mean of both emotion ratings during manic states is more positive and activated compared to the corresponding within-subject ratings during depressed states. Significance was determined by a t-test with $p < 0.01$. This provides evidence that emotion behavior may be effective for predicting mood states.\vspace{-0.5mm}
    \item For almost all subjects, activation and valence are significantly positively correlated with YMRS and significantly negatively correlated with HamD ($p < 0.01$). This supports the hypothesis that heightened mood states come with heightened emotions.\vspace{-0.5mm}
    \item Even after normalizing each subject by his/her euthymic segments, the distribution of emotion ratings between subjects is significantly different (using a one-way ANOVA with $p < 0.01$). We also used a Tukey-Kramer posthoc test of the 66 possible pairwise subject comparisons. Activation was found to be significantly different in 51 cases and valence was found to be significantly different in 48 cases ($p < 0.01$).\vspace{-0.5mm}
    \item Our experiments did not show a correlation between the within-call variance of emotion ratings and mood states.
\end{itemize}

\vspace{-2mm}
\section{Conclusion}
In this work, we present the PRIORI Emotion Dataset - a natural dataset of emotional speech passively-recorded from patients with BD. The dataset is unique in that it has a high proportion of emotional segments of speech and is the only \emph{in the wild} telephonic dataset, annotated for emotion. The dataset contains more than \emph{25} hours of speech (\emph{13,611} segments) with the average of \emph{3.83} labels per segment. We train a CNN model and show that it is possible to accurately extract activation ratings from unstructured speech. We achieve a PCC of \emph{0.712} and \emph{0.405} for activation and valence, respectively. We perform exploratory analysis to show how these predicted emotion labels, normalized using subject's euthymic baseline, correlate to YMRS and HamD values. We find that mean of both emotion ratings during manic states is significantly higher than the mean of the corresponding ratings during depressed states.

The annotation process is ongoing.   After annotation concludes, this dataset will be released to the community due to its potential to impact the field of affective computing. Future work includes implementing subject-dependent models for emotion recognition so that we can directly predict the ratings instead of normalizing them post-hoc. This would help in building subject-dependent mood recognition. 

\vspace{-2mm}
\section{Acknowledgements}
Toyota Research Institute (``TRI'') provided funds to assist the authors with their research but this article solely reflects the opinions and conclusions of its authors and not TRI or any other Toyota entity. This work was supported by the National Science Foundation (CAREER-1651740). NIMH R34MH100404, the Heinz C Prechter Bipolar Research Fund and the Richard Tam Foundation at the University of Michigan. The authors would also like to thank Holli Bertram, Hu Zhaoxian and Beth Semel for their help in collecting PRIORI Emotion Dataset.

\bibliographystyle{IEEEtran.bst}

\bibliography{mybib.bib}

\end{document}